\documentclass{SciPost}

\hypersetup{
    colorlinks,
    linkcolor={red!50!black},
    citecolor={blue!50!black},
    urlcolor={blue!80!black}
}

\usepackage[bitstream-charter]{mathdesign}
\usepackage{wrapfig} 
\DeclareSymbolFont{usualmathcal}{OMS}{cmsy}{m}{n}
\DeclareSymbolFontAlphabet{\mathcal}{usualmathcal}

\fancypagestyle{SPstyle}{
\fancyhf{}
\lhead{\colorbox{scipostdeepblue}{\bf \color{white} ~SciPost Physics Proceedings }}
\rhead{{\bf \color{scipostdeepblue} ~Submission }}

\fancyfoot[C]{\textbf{\thepage}}
}

\begin{document}

\pagestyle{SPstyle}

\begin{center}{\Large \textbf{\color{scipostdeepblue}{
Particle Identification with MLPs and PINNs Using HADES Data\\
}}}\end{center}

\begin{center}\textbf{
Marvin Kohls\textsuperscript{1$\star$}
}\end{center}

\begin{center}
{\bf 1} GSI Helmholtzzentrum für Schwerionenforschung GmbH
\\[\baselineskip]
$\star$ \href{mailto:m.kohls@gsi.de}{\small m.kohls@gsi.de}
\end{center}

\definecolor{palegray}{gray}{0.95}
\begin{center}
\colorbox{palegray}{
  \begin{tabular}{rr}
  \begin{minipage}{0.37\textwidth}
    \includegraphics[width=60mm]{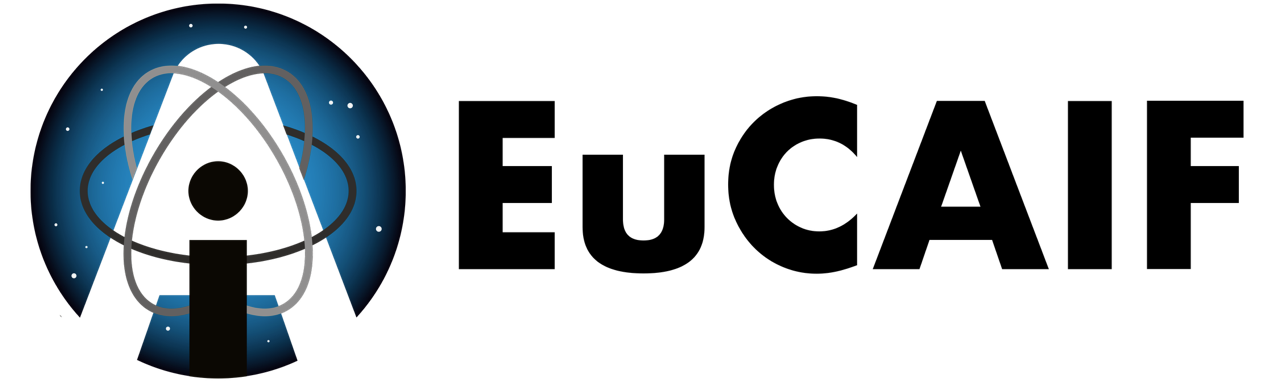}
  \end{minipage}
  &
  \begin{minipage}{0.5\textwidth}
    \vspace{5pt}
    \vspace{0.5\baselineskip} 
    \begin{center} \hspace{5pt}
    {\it The 2nd European AI for Fundamental \\Physics Conference (EuCAIFCon2025)} \\
    {\it Cagliari, Sardinia, 16-20 June 2025
    }
    \vspace{0.5\baselineskip} 
    \vspace{5pt}
    \end{center}
    
  \end{minipage}
\end{tabular}
}
\end{center}

\section*{\color{scipostdeepblue}{Abstract}}
\textbf{\boldmath{%
In experimental nuclear and particle physics, the extraction of high-purity samples of rare events critically depends on the efficiency and accuracy of particle identification (PID). In this work, we present a PID method applied to HADES data at the level of fully reconstructed particle track candidates. The results demonstrate a significant improvement in PID performance compared to conventional techniques, highlighting the potential of physics-informed neural networks as a powerful tool for future data analyses.
}}

\vspace{\baselineskip}

\noindent\textcolor{white!90!black}{%
\fbox{\parbox{0.975\linewidth}{%
\textcolor{white!40!black}{\begin{tabular}{lr}%
  \begin{minipage}{0.6\textwidth}%
    {\small Copyright attribution to authors. \newline
    This work is a submission to SciPost Phys. Proc. \newline
    License information to appear upon publication. \newline
    Publication information to appear upon publication.}
  \end{minipage} & \begin{minipage}{0.4\textwidth}
    {\small Received Date \newline Accepted Date \newline Published Date}%
  \end{minipage}
\end{tabular}}
}}
}




\section{The HADES Experiment}
\label{sec:HADES}
\begin{figure}[h]
    \centering
    \includegraphics[width=10.0cm]{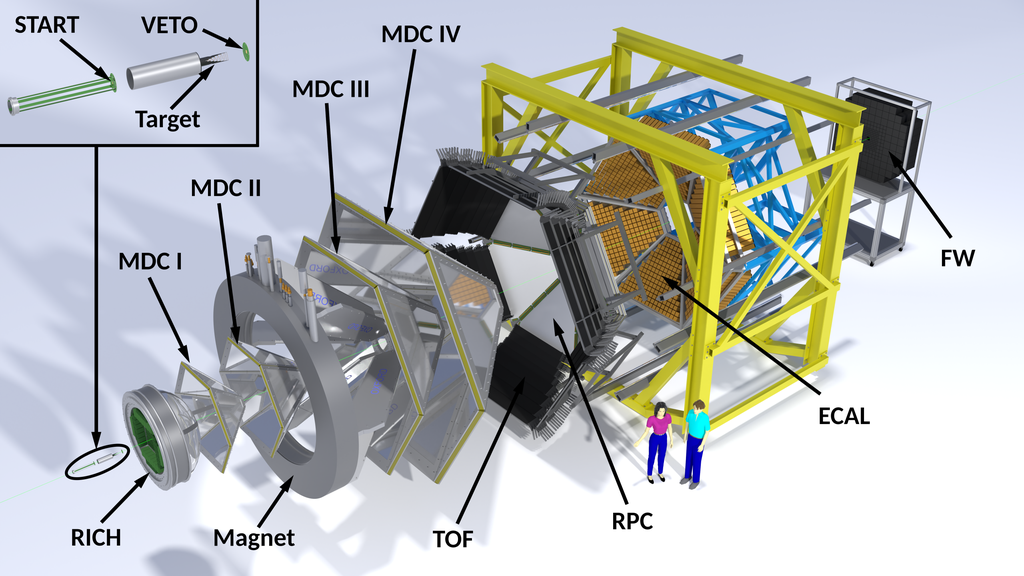}
    \caption{Exploded view of he HADES experimental setup.} 
    \label{fig:HADES}
\end{figure}

The High-Acceptance Di-Electron Spectrometer (HADES) is a fixed-target experiment located at the SIS18 accelerator facility at the GSI Helmholtzzentrum für Schwerionenforschung in Darmstadt, Germany. 
One of the physics goals of HADES is to investigate dense nuclear matter under conditions similar to those in neutron star mergers \cite{HADES:2019auv}. Moreover HADES is well-suited to perform studies related to hadron physics: the few-GeV energy regime offers a unique laboratory for studying nucleon resonances in the second and third resonance regions and for exploring strangeness production at its energetic threshold. A distinctive feature of HADES at SIS18 is the dedicated pion beam programme \cite{HADES:2024piqcd}.
The detector is designed as a magnet spectrometer dedicated to the reconstruction of charged particles, particularly electrons and positrons. Figure~\ref{fig:HADES} shows an exploded view of the experimental setup.
A beam particle first traverses the START detector, generating a $t_0$ signal, which is subsequently used as a reference time. If no reaction occurs between the beam particle and the target, the VETO detector is triggered, and the event is discarded. If an interaction takes place, the forward-boosted particles enter the main HADES detector, which provides nearly full $360^{\circ}$ azimuthal angle and a polar angle coverage from $18^{\circ}$ to $85^{\circ}$.\\
The Ring Imaging Cherenkov Detector (RICH) is a gas-filled detector for identifying electron–positron pairs via their Cherenkov light cones, which are recorded by an array of multi-anode photomultipliers. The gas-filled Multiwire Drift Chambers (MDCs) are used for tracking and energy loss measurements of charged hadrons. The segments of the inner chambers are used to reconstruct the initial point of interaction between the beam and target particles, as well as for tracking.\\
Inner and outer track segments are iteratively combined via cubic spline fits, followed by a Runge–Kutta algorithm that incorporates the Lorentz force experienced by the hadron in the localised magnetic field.\\
Reconstruction of the time of flight (ToF) of the charged hadron is done via a corresponding signal within the META ToF-wall, consisting of Resistive Plate Chambers (RPC) at polar angles below 45$^{\circ}$ and scintillator bars at larger angles, in combination with the $t_0$ signal from the START detector. Matching a Runge–Kutta-fitted track with the corresponding META hit yields the final particle track candidate used for the PID studies presented in this work.\\
\section{Particle Identification}
As introduced above, PID can be performed on the basis of four fundamental observables: particle track length, particle track polarity (direction of bending in the magnetic field), time of flight and integrated specific energy loss. Additionally, $\chi^2$ values for the Runge-Kutta fit as well as for the matching between reconstructed track and META hit can be used as an additional criterion for the quality of the used tracks.
Conventionally, PID for individual charged hadrons in HADES is performed via the correlation in $\beta$ as a function of p/q, combined with a selection on the specific energy loss as a function of p/q.\\
To perform multi-differential analyses with the selected particles, hadron yields are typically extracted from binned distributions in rapidity ($y$)—representing the longitudinal momentum component along the beam axis—as a function of either transverse mass ($m_{\text{t}}$) or transverse momentum ($p_{\text{t}}$), which represent the mass and momentum in the plane perpendicular to the beam axis, respectively. The yields for a given hadron in each $(p_{\text{t}},y)$ bin are obtained from the corresponding mass spectra. To this end, it is desirable to achieve a high purity sample of the particle which is to be further analysed.
\subsection{Short-Lived Particles}
For the reconstruction of short-lived particles via their measured daughter-particles, additional topological constraints can be applied to enhance the signal otherwise dominated by combinatorial background. This holds true, in particular, for particles decaying at least several cm off the initial interaction vertex. These so-called off-vertex decays can be well constrained via the separation of their daughter-particles' point of origin from the other particles of the event coming from the primary event vertex. Using loose preselection criteria, Multi-Layer-Perceptrons (MLPs) have been used successfully in \cite{Spies22} to optimise the selection criteria.
\subsection{Charged Hadron Identification with Neural Networks}
\begin{figure}[h]
    \centering
    \includegraphics[width=15.0cm]{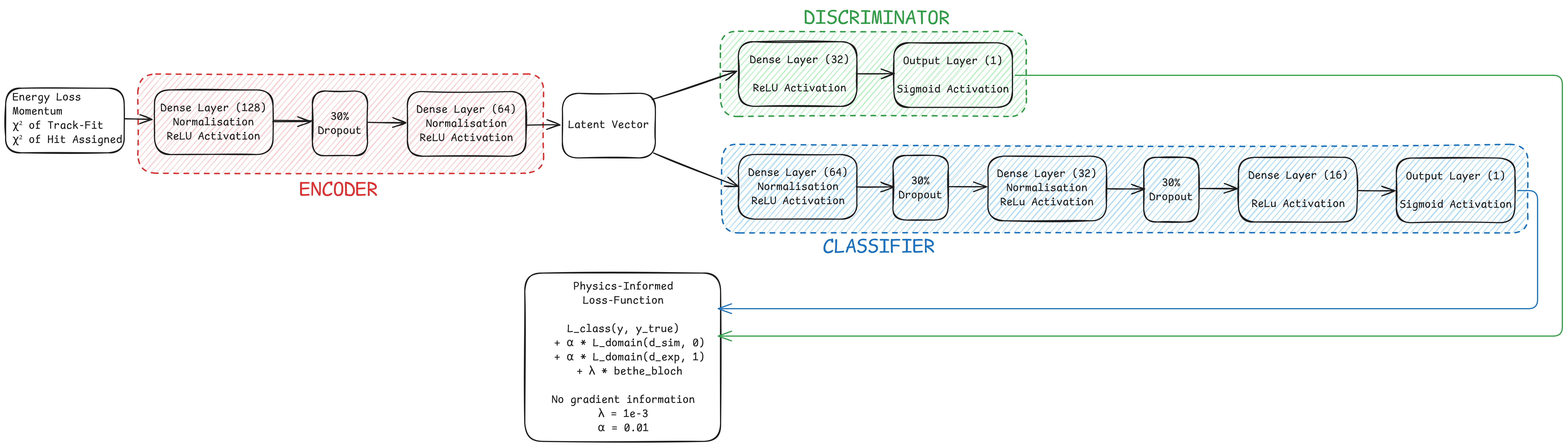}
    \caption{Setup of the physics informed neural network for particle identification.} 
    \label{fig:PIDANN}
\end{figure}

Building on the work on Domain Adversarial Neural Networks (DANNs) presented in~\cite{Ganin:2015usz} and the conceptual framework of Physics-Informed Neural Networks (PINNs), e.g. in \cite{RAISSI2019686}, which incorporate first-principle physics concepts into the loss function, this study extends the application of Artificial Neural Networks (ANNs) to the identification of single charged hadrons based on their features.\\
To achieve this, the neural network was designed in the way depicted in figure \ref{fig:PIDANN}.\\

Specific energy loss and momentum are two available observables reconstructed independently from one another. Therefore, together with track fit quality parameters, they are included as input features for ANN-based PID.\\
The training dataset consists of a split into simulated, labelled data (50$\%$) and experimental, unlabelled data (50$\%$). These are mapped by an encoder network into a feature space represented by latent vectors. These latent vectors are, on one hand, domain-classified by a discriminator according to their origin (experimental or simulated). This allows the network to learn and compensate for discrepancies between simulated and experimental data. This prevents over-reliance on simulation-specific distributions, which could otherwise bias the classification.
On the other hand, a classifier predicts the particle type present in the event based on its feature-space representation. This classification may be binary or based on a predefined set of labels.\\
The output of both networks is combined into a custom loss-function which uses the momentum information from the original particle to calculate the theoretically expected value for the specific energy loss of the desired particle at the given momentum and, subsequently, calculates a loss $L$ according to 
\begin{equation}
    L = L_{\text{class}}(y,y_{\text{true}}) + \alpha \cdot L_{\text{domain}}(d_{\text{sim}},0) + \alpha \cdot L_{\text{domain}}(d_{\text{exp}},1) + \lambda \cdot L_{\text{Bethe-Bloch}}\left( \frac{\mathrm{d}E}{\mathrm{d}x}, \frac{\mathrm{d}E}{\mathrm{d}x}_{\text{theory}} \right)
\end{equation}
where $L_{\text{Bethe-Bloch}}$ implements the Bethe-Bloch formula for charged Kaons ($m_K=493.677$ MeV/$c^2$) in the MDC gas mixture (70\% Ar, 30\% CO$_2$)\footnote{The physics penalty is implemented as a custom Keras layer that calculates the theoretical d$E$/d$x$ from momentum and adds the MSE between measured and expected values with $\lambda=10^{-3}$}.
After loss evaluation, the gradient is backpropagated to update the classifier weights for the next training iteration.\\
Correspondingly, we focus on the reconstruction of $K^+$-mesons without applying preselection criteria. These mesons are rare and only produced once in every 100th event in the specific presented case of Ag(1.58 A GeV)+Ag collisions.\\
\begin{wrapfigure}[14]{r}{0.45\textwidth} 
    \centering
    \includegraphics[width=0.43\textwidth]{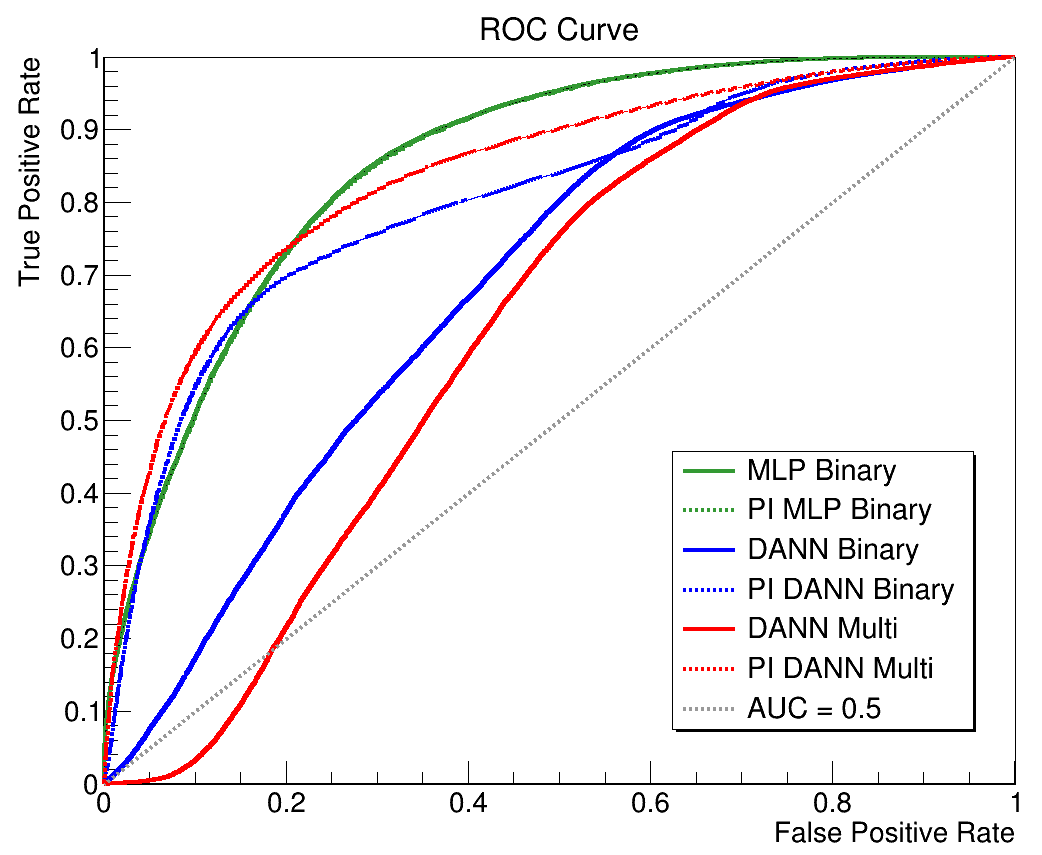}
    \caption[ROC Curve]{Performance for different ANN setups in identifying $K^{+}$ in Ag(1.58 A GeV)+Ag data. PI stands for "Physics Informed" neural networks.} 
    \label{fig:ROC}
\end{wrapfigure}
\section{Results}
We evaluated six different neural network setups: MLPs with and without physics information, DANNs with binary classification with and without physics information, and DANNs with multi-label classification with and without physics information. The performance is summarised by the receiver operator curves (ROCs) in Figure~\ref{fig:ROC}. Incorporating physics information into the MLP loss function does not yield a noticeable change. In contrast, adding physics information to the loss function of both binary and multi-label DANN classifiers substantially improves performance, resulting in comparable area-under-the-curve (AUC) values (approximately 0.82 for DANNs and 0.85 for MLPs).
\section{Discussion and Conclusion}
The apparent insensitivity of the MLP to the additional physics term in the loss function suggests that it learns predominantly from simulation-specific features, which can introduce bias when applied to experimental data. Consequently, MLPs may be suboptimal for the PID tasks considered here. Because the DANN operates on latent representations in feature space, future work should incorporate additional observables to further constrain the domain alignment and classification. Moreover, the current loss formulation assumes that the mean of the Bethe–Bloch curve corresponds to the theoretically calculated value. Since the measured specific energy loss follows a Landau-like distribution with asymmetric tails, the physics term should be revised to reflect the appropriate statistical fluctuations.\\
In summary, the results indicate that adding physics information to the DANN loss function can substantially increase the purity of the selected sample. This approach will be further explored and extended in future studies.
\section*{Acknowledgements}
This research was supported by GSI Helmholtzzentrum für Schwerionenforschung GmbH. The author also gratefully acknowledges the HADES Collaboration for providing the data, and Simon Spies and Waleed Esmail, whose earlier work on neural network–based analyses of HADES data formed an inspirational foundation for this study. All figures are classified as original work.

\bibliography{SciPost_Example_BiBTeX_File}


\end{document}